\documentclass[11pt]{article} 
\usepackage{epsf,axodraw} 
\newcommand{\bea}{\begin{eqnarray}}
\newcommand{\eea}{\end{eqnarray}}
\begin{document}

\newcommand{\lsim}
{{\;\raise0.3ex\hbox{$<$\kern-0.75em\raise-1.1ex\hbox{$\sim$}}\;}}
\newcommand{\gsim}
{{\;\raise0.3ex\hbox{$>$\kern-0.75em\raise-1.1ex\hbox{$\sim$}}\;}}

\begin{flushright}
{\large 
{HRI-04-04-001}\\ 
{HIP-2004-11/TH}\\
{ROME1-1372-2004} \\
{\Large \tt hep-ph/}\\
}
\end{flushright}
\vspace{0.1cm}

\begin{center}
{\LARGE\bf Invisible Higgs in theories of large extra dimensions}
\\[15mm]
{\bf Anindya Datta$^a$, Katri Huitu$^{b,c}$, Jari Laamanen$^b$, 
and Biswarup Mukhopadhyaya$^d$}
\\[4mm]
$^a$INFN, Sezione di Roma, Universita La Sapienza\\
P. le A. Moro 2, Rome I-00185, Italy\\[4mm]
$^b$Helsinki Institute of Physics \\
P.O.Box 64, FIN-00014 University of  Helsinki, Finland \\[4mm]
$^c$High Energy Physics Division, Department of Physical Sciences,
\\
P.O.Box 64, FIN-00014 University of  Helsinki, Finland \\[4mm]
$^d$Harish Chandra Research Institute, Chhatnag Road, Jhusi\\
Allahabad - 211 019, India\\[7mm]
\date{}
\end{center}

\begin{abstract}

We discuss the possibility of detecting a Higgs boson in future
collider experiments if large extra dimensions are realized in nature.
In such a case, the Higgs boson can decay invisibly by oscillating
into a graviscalar Kaluza-Klein (KK) tower. We show that the search
for such a Higgs at an $e^+ e^-$ linear collider entails more
complications than are usually thought of in relation to an invisibly
decaying Higgs, the main sources of such complications being the
simultaneous presence of a continuum graviton production and the
broadening of the Higgs peak. We discuss possible ways of overcoming
such difficulties, and conclude that the detection of such a Higgs
boson might still be a problem beyond the mass range of 250-300 GeV.
\end{abstract}

\vskip 7mm

\section{Introduction} 
     \label{sect:intro}
Confirming the Higgs mechanism as the underlying principle 
of electroweak symmetry breaking is one of the 
main goals of upcoming accelerators.
It is expected that the Large Hadron Collider (LHC) 
will detect the particle responsible for the symmetry breaking
in the Glashow-Salam-Weinberg model, namely, the Higgs $H$.
However, for studying its properties in detail, 
a linear electron-positron collider will probably be needed.

The strategy for Higgs search depends on the decay branching fractions of
Higgs to different channels. This is where the likely modification of
Higgs signals due to physics beyond the Standard Model (SM) comes up
for consideration, since such new physics is very likely to affect the
Higgs interactions rates and its consequent decay products in various
ways. One somewhat bizarre (but by no means inconceivable) possibility
in this context is that a Higgs can have a substantial branching
ratio for decay into invisible final states. This possibility has been
underlined in a number of well-motivated theoretical options
\cite{inv,invLHC,EZ,susy,majoron,dpinv,GRW,neutrino}.

It is expected in some of these options that such invisible decays
will dominate, making the Higgs boson difficult to identify
in collider experiments. It is therefore a rather interesting subject 
of investigations as to the nature and extent of invisibility
acquired by Higgs, and how it can be related to specific aspects
of the theories concerned, whether  it be 
supersymmetry \cite{susy}, majoron models \cite{majoron,dpinv} or models
with extra compact spacelike dimensions \cite{GRW,neutrino}.
Our interest in this paper is to study the invisible decay modes
suggested in models with extra dimensions.

Theories of the aforementioned type have been popular in recent times,
because (a) they offer a rather appealing solution to the hierarchy
problem, and (b) they entail the prospect of testing gravitational
effects in TeV-scale experiments..  Broadly two such types of models
have been studied so far, namely, the Arkani-Hamed-Dimopoulos-Dvali
(ADD) \cite{ADD} and Randall-Sundrum (RS) \cite{RS} types. Both of
them envisage extra compact spacelike dimensions, with gravity
propagating in the `bulk', while all the SM fields are confined to
(3+1) dimensional slices or `branes' in the minimal versions of both
types. In ADD-type models, one has a factorizable geometry, where the
projection on the brane leads to a continuum of scalar and tensor
graviton states, whose cumulative effect gives new contributions at
the TeV scale, thus providing a natural cut-off to SM physics. In
RS-type models, on the other hand, a non-factorizable geometry is
assumed, where the size of the fifth dimension is small and hierarchy
is removed by a `warp' factor with a large negative exponent, which
scales down large mass parameters on the visible brane.  The
discussion in this note is related to the former type of models.

An important feature of these models is that the Higgs boson can mix
with the graviscalars (the tower of scalar states arising out of the
projection of the graviton on the visible brane). Although it is not
usually retained in the minimal model, it is perfectly consistent with
general covariance to augment the four-dimensional effective action
with a term
\bea
S=-\xi\int d^4 x\,\sqrt{-g_{ind}}R(g_{ind})H^{\dagger}H,
\eea
where $H$ is the Higgs doublet, $\xi$ is a dimensionless mixing
parameter, $g_{ind}$, the induced metric on the brane and $R$ is the
Ricci scalar.  As has been shown in \cite{GRW}, this term basically
arises from the fact that the energy-momentum tensor $T_{\mu\nu}$ can
be extended by a term of the type $\xi
(\eta_{\mu\nu}\partial_{\alpha}\partial^{\alpha} -
\partial_{\mu}\partial_{\nu})$ while it is still conserved.
Once electroweak symmetry is broken, the coupling of the trace of 
the additional part to the graviscalars leads to a mixing between
the physical Higgs field (h) and each member of the graviscalar
tower. One can parametrize such mixing by the following term 
in the Lagrangian:

\begin{equation}
{\cal L}_{mix} = \frac{1}{M_P} m_{mix}^3h\sum_n S_n,  
\label{mix}
\end{equation}

\noindent
where, $m_{mix}^3 = 2 \kappa \xi v m_h^2$, $M_P$ is the reduced Planck
mass, and $v$ is the Higgs vacuum
expectation value. $\kappa$ can be expressed in terms of the number of
extra dimensions: 
\begin{equation}
\kappa \equiv \sqrt{\frac{3 (\delta -1)}{\delta +2}}
\end{equation}

\noindent 
$\delta$ being the number of extra compact spacelike dimensions.

It should be mentioned here that the effects of extra dimensions on the
Higgs decay modes are different in the case of large or small
extra dimensions.  This is due to the different spacing of the
KK-towers. In the case of the small extra dimensions the spacing is
large, and the observable effect is expected to come from the
nontrivial mixing of the Higgs boson with the single graviscalar in
the model, called radion \cite{RS}. The consequences of
radion-Higgs mixing have already been explored in the literature
\cite{GRW,rHmixing}. 
In the case of the large extra
dimensions the major effects are due to the closely spaced KK-levels.
These lead to the possibly effective `invisible decays' of the Higgs boson, 
via oscillation into one or the other state belonging to the 
quasi-continuous tower of graviscalars \cite{GRW}. The consequence of 
this strange phenomenon in the context of collider experiments is
our concern in this note.

In recent years the invisible decays of the Higgs
boson have been under scrutiny, and methods have been developed to
detect the invisible Higgs.  These are based on tagging of some other
particle than the decay products of the Higgs.
Mass limits for a Higgs boson decaying dominantly to invisible
particles have also been obtained by the LEP experiments \cite{ALDO}.
It is assumed that the Higgs boson is produced in association 
with a $Z$-boson, which
decays either to charged leptons or hadrons, and a constraint that 
the decay products are consistent with the $Z$ mass is applied.
Then the mass limit of $m_H>114.4$ GeV can be established.
What we wish to emphasize here is that {\em a scenario, where
invisibility is induced by scalar-graviscalar mixing, brings in
some additional complications in detecting Higgs signals. Some
ways out of such complications are suggested here}.

In section 2 we present the branching ratios for invisible decay
in terms of the fundamental parameters of theory. Different aspects
of the ensuing Higgs signals in a high-energy $e^+ e^-$ collider
are discussed in section 3. We summarise and conclude in section 4.


\section{Higgs invisible decays with scalar-graviscalar mixing}

The usual way to handle a case of mixing (as defined in 
Eq.~(\ref{mix})) is to diagonalise the mass matrix and 
define the physical scalar eigenstates. However,
due to the presence of an infinite number of members of the graviscalar
tower, it is technically difficult to tackle this problem in this way.

Instead, one proceeds by considering the Higgs propagator in the flavour
basis itself and incorporating all the insertions induced by the mixing term. 
This requires one to integrate over a tower of quasi-continuous states.
As has been shown in \cite{GRW}, the provision of thus having a large
number of real intermediate states inserted leads to the development
of an imaginary term in the propagator. This imaginary part
can be interpreted as an effective decay width entering into
the propagator after the fashion of the Breit-Wigner scheme 
\cite{GRW}:\footnote{Our calculation of the width agrees with \cite{BDGW}
where the width is twice the value in \cite{GRW}.} 
\bea
\Gamma_G=2\pi\kappa^2\xi^2v^2\frac{m_h^{1+\delta}}{M_D^{2+\delta}}
S_{\delta -1},
\label{osc}
\eea
where $M_D$ is the $(4+\delta)$-dimensional Planck scale (also 
called the string 
scale) and  $S_{\delta -1}$ is the surface of a unit-radius sphere in 
$\delta$ dimensions, given by
\begin{equation}
S_{\delta -1}~=~{\frac{2\pi^{\delta/2}}{\Gamma(\delta/2)}}
\end{equation}
This invisibility can be understood qualitatively in the following
way.  Due to the presence of the mixing terms, a Higgs boson, once produced,
has a finite probability (proportional to $\xi^2$)
of oscillating into a quasi-continuous multitude of invisible
states, corresponding to the graviscalar tower. Though the mixing between
the Higgs and any one of the graviscalar is the same, this transition
would be favoured when masses of the Higgs and the corresponding
graviscalar are close to each other. The mixing is suppressed by
$M_P$. But summing the oscillation probabilities over the tower
can make the {\em resulting probability for the Higgs going into one of
many graviscalars quite
big}. Couplings of each graviscalar to ordinary matter being 
suppressed by Planck mass, any of the states the Higgs can oscillate into
is practically invisible. Thus, the Higgs, once transformed into a graviscalar,
becomes invisible and this is reflected in the `invisible' decay width 
developed by the propagator. 

\begin{figure}[t]
\begin{center}
\hspace{-1in}
\centerline{\hspace*{3em}
\epsfxsize=7cm\epsfysize=8cm
                     \epsfbox{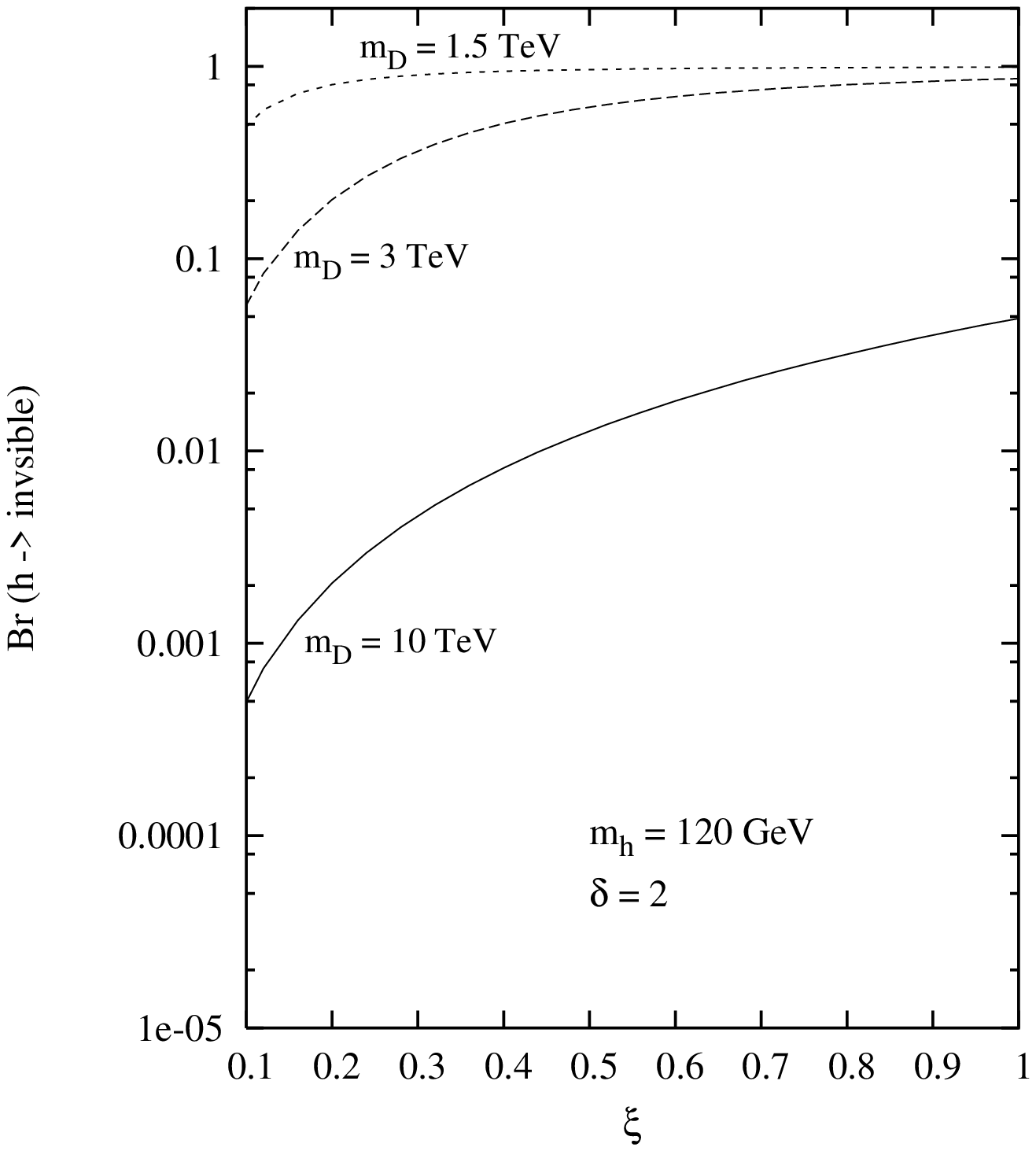}
\epsfxsize=7cm\epsfysize=8cm
                     \epsfbox{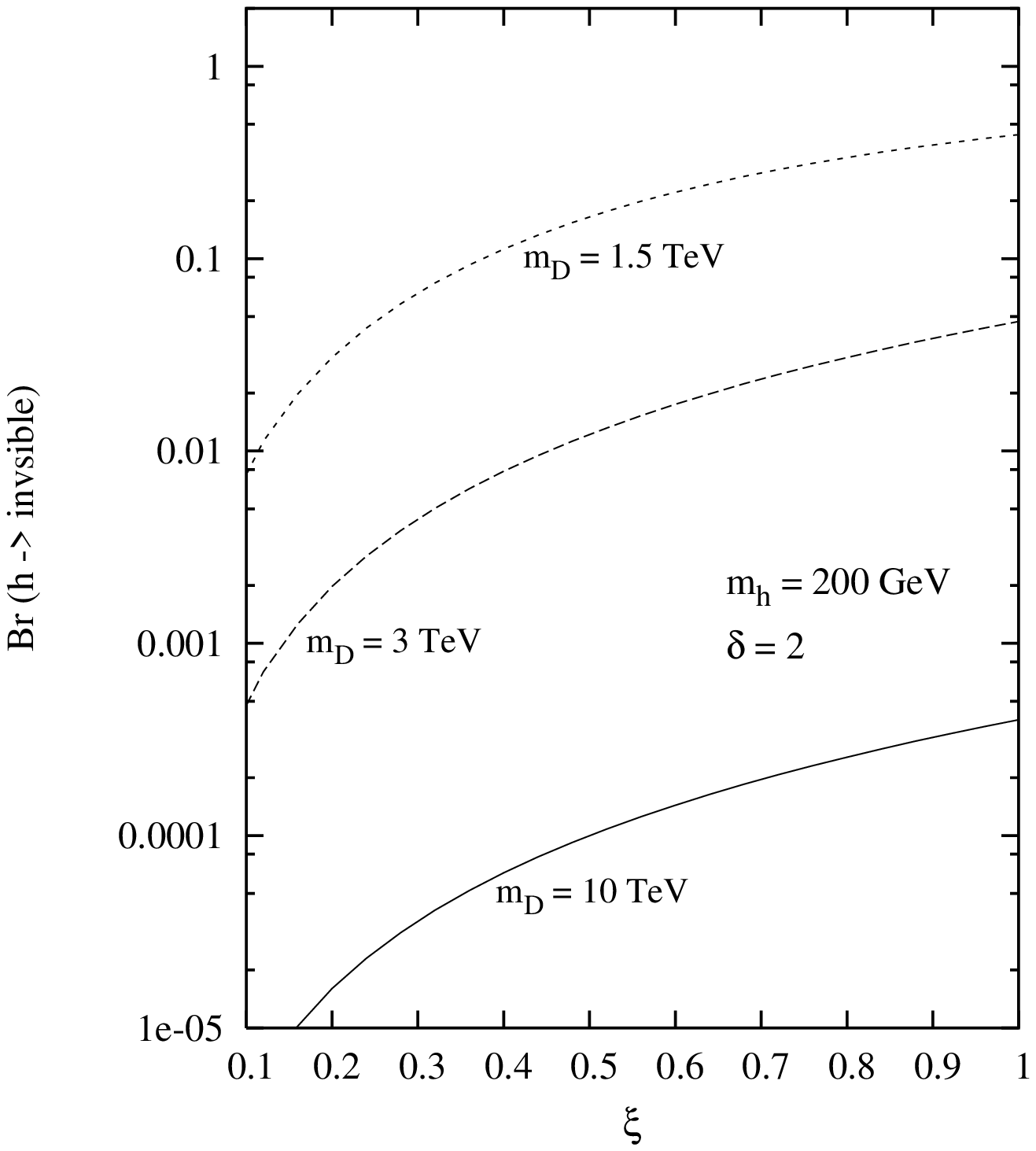}
}
\end{center}

\caption{\label{invBR}The invisible decay branching ratio of the Higgs boson 
as a function of the mixing parameter for two representative values of
Higgs mass and for number of extra dimension, $\delta = 2$. }

\end{figure}

Assuming that the graviton KK tower is the only source of the
invisible width in the model, we have plotted in Fig.~\ref{invBR} the
invisible branching ratio as a function of the mixing parameter. Two
masses for the Higgs boson, namely, $m_h=120$ GeV, and $m_h=200$ GeV,
have been used. The plots have been made for $M_D=1.5$, 3, and 10 TeV
and for $\delta$=2. In Table~\ref{tabwid}, total Higgs boson widths
are presented for two values of $m_D$, namely 1.5 and 3 TeV; in
presence of Higgs-graviscalar mixing.  For $m_h = 120$ GeV, width of
the SM Higgs boson is of the order of $10^{-3}$ GeV. With the on-set of
mixing term, the width has become $0.62$ GeV for a more favorable case
($m_D =$ 1.5 TeV, $\delta =$2), which is evident from the table. The
decrement of the width with increasing $\delta$ can be explained by
the steady decrease of number density of graviton states in which the
Higgs could oscillate. A closer look to the numbers in the second row 
(for a Higgs mass of 200 GeV), reveals the importance of $WW, ZZ$ decay
channels for Higgs and the comparative strength of these with
invisible channel. Decay width of a 200 GeV SM Higgs boson is 
nearly 2 GeV.  At this mass we see that invisible width is 
comparable with  visible channels.

\begin{table}
\begin{center}
\begin{tabular}{|c|c|c|c|c|c|c|}
\hline
\multicolumn{7}{|c|}{Decay width (GeV) of the Higgs boson} \\
\hline
& \multicolumn{3}{|c}{$m_D$ = 1.5 TeV} &
\multicolumn{3}{|c|}{$m_D$ = 3 TeV}  \\
 \cline{2-7} {} & \multicolumn{1}{|c}{$\delta$ = 2} &
\multicolumn{1}{|c|}{$\delta$ = 3} & \multicolumn{1}{|c}{$\delta$ = 4}
 & \multicolumn{1}{|c}{$\delta$ = 2} &
\multicolumn{1}{|c|}{$\delta$ = 3} & \multicolumn{1}{|c|}{$\delta$ = 4}\\ \hline
\multicolumn{1}{|c|}{$m_h =$ 120 GeV}
& \multicolumn{1}{|c}{0.62} & \multicolumn{1}{|c|}{0.163} &
\multicolumn{1}{|c}{0.032} & \multicolumn{1}{|c|}{0.044} &
\multicolumn{1}{|c}{0.016} & \multicolumn{1}{|c|}{0.011}  \\ 
\hline
\multicolumn{1}{|c|}{$m_h =$ 200 GeV}
& \multicolumn{1}{|c}{6.445} & \multicolumn{1}{|c|}{4.814} &
\multicolumn{1}{|c}{3.918} & \multicolumn{1}{|c|}{3.778} &
\multicolumn{1}{|c}{3.638} & \multicolumn{1}{|c|}{3.608}  \\ 
\hline

\end{tabular}
\end{center}
\caption{\label{tabwid}Total decay width of Higgs boson in presence
of Higgs-graviscalar mixing with mixing parameter $\xi = 1$.}
\end{table}

The message emerging from the above discussion is that this
effective invisible decay width grows as $m_h ^3$ for $\delta = 2$. 
This implies that even for ($m_h < 2 m_W$) total Higgs decay
width can be considerably larger than the Standard Model width
which in this range is dominated by the decay to fermion pairs and is
proportional to $m_H$. As a consequence, even for a light
Higgs boson, Higgs resonance may not be very sharp. This fact has to be 
taken into account when one attempts to reconstruct such a Higgs
through the recoil invariant mass peak at a linear electron-positron
collider.

Invisible decay of Higgs boson is also possible in  models with
right-handed neutrinos propagating in the bulk \cite{neutrino}. However,
in this article we will confine ourselves to the case of Higgs-graviscalar 
mixing only.

\section{$e^+e^-$ colliders}

It is evident from the figures presented in the previous section that
the Higgs can have a very large invisible branching ratio in the
scenario considered here. The identification of such a Higgs at hadron
colliders is in general a difficult task; associated productions such
as $WH$ has been studied at Tevatron and LHC \cite{dpinv} only in the
context of four dimensional models. The gauge boson fusion (GBF)
channel may be interesting at the LHC, but is again plagued with
complications, since ADD gravitons can also be produced by the same
mechanism. Thus, unless a full and reliable calculation of graviton
production is available, the clue on whether an invisible Higgs is
produced alongside may be lost from our sight.  

We consider the situation at an $e^+e^-$ machine, where the 
identification of the recoil mass peak against the $Z$-boson in the
Bjorken process is widely known to be a reliable method of
detecting the Higgs boson.  We therefore concentrate on this process, 
also known as  the Higgs-strahlung process, at a linear
collider with center of mass energy of 1 TeV,
\bea
e^+e^-\rightarrow Z(\rightarrow \mu^+\mu^- )h(\rightarrow {\rm inv}).
\eea
The final state we are interested in comprises of a $\mu ^+ \mu ^-$
pair with missing energy/momentum (corresponding to one or the other  
of the graviscalars). Due to a sizable Higgs width even
for light Higgs bosons, we do not use the narrow width approximation,
but rather treat 
the Higgs boson as a propagator (inclusive of an invisible width)
while calculating the cross-section.
We have also taken into account the direct graviscalar production
alongside.

At the center-of-mass energies we are concentrating on, the production 
of Higgs boson in $ZZ$- and $WW$-fusions is also important. 
However, since we are interested in the
invisible decay of the Higgs boson, the $WW$-fusion channel is not suitable
for our purpose, as in the final state we are left with an invisibly 
decaying Higgs with two neutrinos. 
A closer look at the Higgs production cross-sections \cite{zerwas} in 
different channels at the $e^+ e^-$  collision reveals that the rate for 
$H e^+ e^-$ in the Higgs-strahlung and $ZZ$ fusion channels may be
comparable for low Higgs masses and center of mass energies higher
than 500 GeV. Thus 
the Higgs/graviscalar production in $ZZ$-fusion channel would also
contain in the final state a charged lepton pair with an invisibly 
decaying Higgs.
The gravitensor production rate should be fully estimated
in the same channel. 
In the present work, we will report only on the Higgs-strahlung process,
which already shows the relative importance of gravitensor and 
Higgs production, as well as the correlation of 'signal' and
'background' in particular examples of invisible Higgs in 
extra dimensional models. That is why we have chosen to illustrate 
our main point by focusing on the $\mu^{+} \mu^{-}$ final state.

\begin{figure}[ht]
{
\unitlength=1.0 pt
\SetScale{1.0}
\SetWidth{0.7}      
\scriptsize    

\begin{minipage}[t]{7.5cm}
\begin{picture}(95,99)(0,0)
\Text(15.0,90.0)[r]{$e^-$}
\Text(15.0,10.0)[r]{$e^+$}
\Text(180.0,75.0)[l]{$Z$}
\Text(100.0,62.0)[l]{$Z$}
\Text(180.0,23.0)[l]{$S_n$ }
\ArrowLine(16.0,90.0)(70.0,50.0)
\ArrowLine(16.0,10.0)(70.0,50.0)
\Photon(70.0,50.0)(135.0,50.0){3}{4}
\Photon(135.0,50.0)(185.0,90.0){3}{4}
\DashLine(135.0,50.0)(185.0,10.0){1.0}
\end{picture}
\end{minipage}
\hfill
\begin{minipage}[t]{7.5cm}
\begin{picture}(95,99)(0,0)
\Text(15.0,90.0)[r]{$e^-$}
\Text(15.0,10.0)[r]{$e^+$}
\Text(180.0,75.0)[l]{$Z$}
\Text(100.0,62.0)[l]{$Z$}
\Text(180.0,21.0)[l]{$S_n$ }
\Text(160.0,28.0)[l]{$\odot$}
\Text(149.0,45.0)[l]{$h$ }
\ArrowLine(16.0,90.0)(70.0,50.0)
\ArrowLine(16.0,10.0)(70.0,50.0)
\Photon(70.0,50.0)(135.0,50.0){3}{4}
\Photon(135.0,50.0)(185.0,90.0){3}{4}
\DashLine(135.0,50.0)(185.0,10.0){1.0}
\end{picture}
\end{minipage}
\hfill
}


\caption{Feynman diagrams contributing to the process 
$e^+ e^- \rightarrow Z S_n$. }
\label{diagrs}
\end{figure}
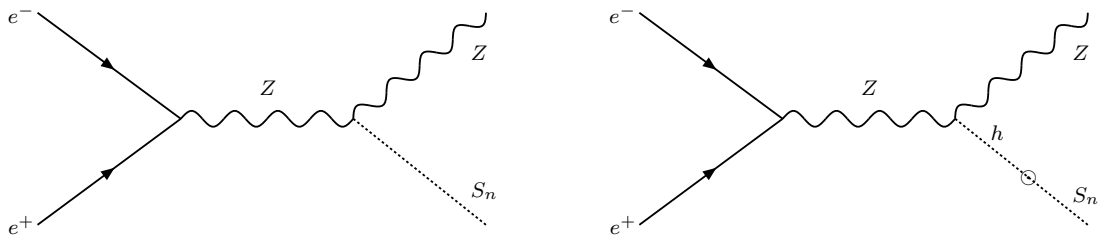

The spin averaged squared matrix element for graviscalar (of mass $m_n$)
production at the $e^+ e^-$ collision via Higgs-strahlung (Feynman 
diagrams are shown in Fig.~\ref{diagrs}) is given by

{\setlength\arraycolsep{2pt}
\begin{eqnarray}
\overline{\sum \vert {\cal M} \vert^2} &=& \frac{g^2}{4 M_P^2}\;
\frac{(c_V^2 + c_A^2)}
{\cos ^2 \theta _W ((s -m_Z^2)^2 + m_Z^2 \Gamma_Z^2) }\;\left[ 2 m_Z^2 s +
2(m_Z^2 - u)\;(m_Z^2 -t) \right] \nonumber \\
&\times&\left\{ \frac{g^2\;m_{mix}^6}
{((m_n^2 -m_h^2)^2 + m_h^2 \Gamma_h^2)} + 
\frac{\kappa^2 (1 - 6 \xi)^2 \cos ^2 \theta _W m_Z^2 }
{9}\right.\nonumber\\ 
&&\left. + \frac{2 g\;m_{mix}^3 \;\kappa \;(1 - 6 \xi)\; (m_n^2 -
  m_h^2) 
\;m_Z}{3 ((m_n^2 -m_h^2)^2 + m_h^2 \Gamma_h^2)}\right\}\!,
\label{matel}
\end{eqnarray}
}
 
\noindent
where $c_V$ and $c_A$ 
are defined as in the coupling of 
an $e^+ e^-$ pair to a $Z$:  $g\;\gamma_\mu \; (c_V + c_A \gamma_5)$. 

The cross-section can be easily obtained from the above after a
straightforward phase space integration.  Finally we have to sum over
the graviscalar tower to have the observable rate. It is, however,
evident from Eq.~(\ref{matel}), that the effect of mixing is most
important for the graviscalar masses close to the Higgs mass.

The important point to note here is that a similar final state can
arise in such a scenarios not only from the Standard Model
contributions ($ZZ/WW$) but also in the production of a $Z$-boson 
with towers of
graviton (spin-2/spin-0). Therefore, a complete calculation on the
prospects of the invisible Higgs signal has to take into account the
continuum graviton production as well.  This is a point that we wish
to emphasize in the present study.

In our calculation, we have included, in addition to the SM processes
$e^+ e^- \longrightarrow ZZ/WW \longrightarrow \mu^+ \mu^- +
\not{\!\!E}$, gravitensor production together with a $Z$-boson, both leading
to identical final states.

We use the following event selection criteria:

\begin{itemize}
\item $ p_T^l > 10$ GeV,
\item reject any $M_{recoil}$ in the mass windows 
of $m_Z \pm 5$ GeV and again $(\sqrt{s} -m_Z) \pm 5$ GeV,   
\end{itemize}

\noindent
where $M_{recoil}$ is defined as the mass of the system recoiling
against the $\mu^+ \mu^-$ pair (which peak at the $Z$-boson mass). The
second criterion is to reject the events originating from SM $ZZ$
production and decay. This kinematic cut also eliminates the signal of
an invisible Higgs, which is very close in mass to a $Z$-boson or of
mass in the vicinity of $(\sqrt{s} -m_Z).$ It is well known that the
Bjorken process is not very useful in searching for a Higgs
boson close in mass to the $Z$. On the other hand, a  part of the
SM background comes from the production of an on-shell $Z$ (subsequently
decaying to a $\mu^+ \mu^-$ pair) along with two neutrinos. A peak in
the recoil mass distribution around $\sqrt{s} - m_Z$ corresponds to
this particular configuration. Our choice of cut on the recoil mass
helps to eliminate this component of background. It is needless to mention 
that, this cut also removes any signal of an invisibly decaying Higgs 
boson around this mass range. But we have investigated that beyond 
a mass of 250 GeV or so, our proposed method is not very effective. 
Even for a 500 GeV $e^+e^-$ machine, the above cut will 
affect the Higgs signal beyond 400 GeV.

\begin{figure}[ht]
\begin{center}
\hspace{-1in}
\centerline{\hspace*{3em}
\epsfxsize=8cm\epsfysize=9cm
                     \epsfbox{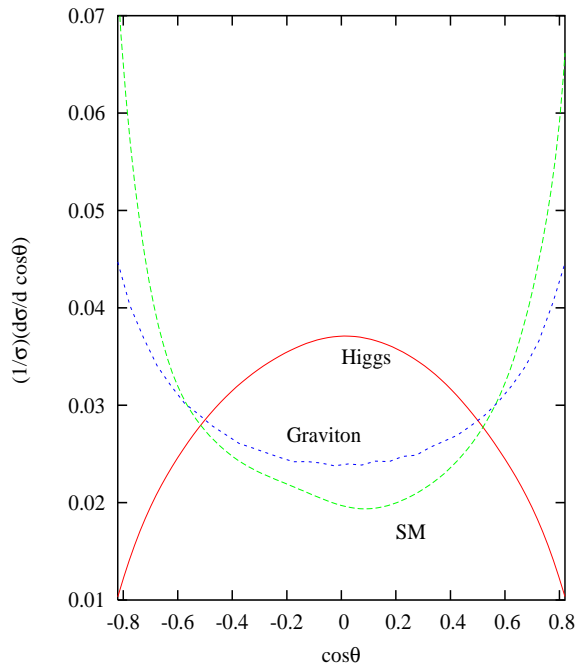}
}
\end{center}
  \caption{ Angular distribution of invisible Higgs as reconstructed from the
 $\mu ^+ \mu^-$ pair. The SM and 
gravitensor distributions are also shown.}
\label{ang_dist}
\end{figure}

We will now concentrate on the possible ways to filter out the Higgs
effects from gravitensor and SM contributions. Though we are dealing
with  invisible decay products in the final state, the environment
of  an $e^+ e^-$ collider enables one to reconstruct 
the invisible system from the
information of the visible decay products. Thus, by measuring the 
four-momenta of the final muons, we can reconstruct the mass of the system
against which this $\mu^+ \mu^-$ are recoiling. In the case of our
interest,  the recoil
mass distribution  peaks at the Higgs mass modulo the Higgs
width and detector resolution. For the gravitensor  production, the
recoil mass distribution is a continuum, reflecting the monotonically
increasing (quasi-) continuous density of graviton states with graviton
mass. The SM contribution, as has been already mentioned, 
has two main channels, namely,
$e^+ e^-\rightarrow ZZ$ and $e^+ e^- \rightarrow W^+ W^-$. 
If now nature has chosen to have large extra dimensions with non-negligible 
Higgs-graviscalar mixing, then the recoil mass distribution in our process
of interest will show a continuum due to gravitensor and SM with
characteristic peaks of Higgs boson superposed on it. The position of
this peak will be determined by the Higgs mass. Now emerges a crucial
issue: the height of this peak. Height of this peak is determined,
apart from other parameters, by the Higgs-graviscalar mixing 
$\xi$, the same quantity which also determines the Higgs width, making
the width large for $\xi~=~O(1)$. As
has already been pointed out, this causes the invisible decay recoil mass
distribution to lose its sharp character even for a Higgs
mass on the order of 120 GeV. A larger width will essentially spread 
out the Higgs contribution to a number of recoil mass bins (centered 
around  the Higgs mass) and thereby weaken the signal. 

One way out of the problem is to use 
angular distributions which are drastically different for Higgs production
as compared to SM and gravitensor production.  If we concentrate on the
distribution of the scattering angle of the $\mu^+ \mu^-$
system, for Higgs production this is
peaked around the central part of the detector in contrast to the
graviton and the SM cases, where these are peaked around the forward
and backward directions with respect to the initial electron beam. 
This can be accounted for by the t-and u- channel electron propagators
in SM and gravitensor production, particularly, when a  
photon (in case of SM) or a very light (nearly massless) graviton
is attached to the electron line. In other words, mass of the final state
particle (other than the $Z$-boson) and also the demand of a minimum
$p_T$ of the 
final state particles, acts as a regulator to this singularity
of cross-section in extreme forward/backward direction.

In Fig.~\ref{ang_dist}, we present the normalised angular distributions 
for above three cases. 
The figure indicates that an angular cut can be used to disentangle the 
two new physics effects from each other, and at the same time to reduce 
the SM backgrounds. The Higgs signal is filtered quite effectively
if we apply the following cut on the relevant angle : 
$ \vert \cos \theta \vert < 0.8$. This selection criteria, 
killing almost 50 \% of the SM and
graviton production, leaves us with a prominent invisible Higgs signal,
at least for a Higgs mass on the lower side.

However, even this angular cut is not of much help  
for higher Higgs masses. This is due to twofold reasons. First, the 
production cross-section itself decreases with rise in the Higgs mass. 
The second and more important reason, however, is 
the sharp increase of Higgs width with Higgs mass. For a 
200 GeV Higgs boson, Higgs width is quite
large due to opening up of $WW, ZZ$ decay channels.  At the same
time the invisible branching ratio can be substantial and even dominant,
as is indicated by Fig. \ref{invBR}. Consequently, the
recoil mass is spread over a larger number of bins, thus flattening out
the peak. This effect is more damaging for our signal the higher the Higgs
mass is.

\begin{figure}[t]
\begin{center}
\hspace{-1in}
\centerline{\hspace*{3em}
\epsfxsize=8cm\epsfysize=9cm
                     \epsfbox{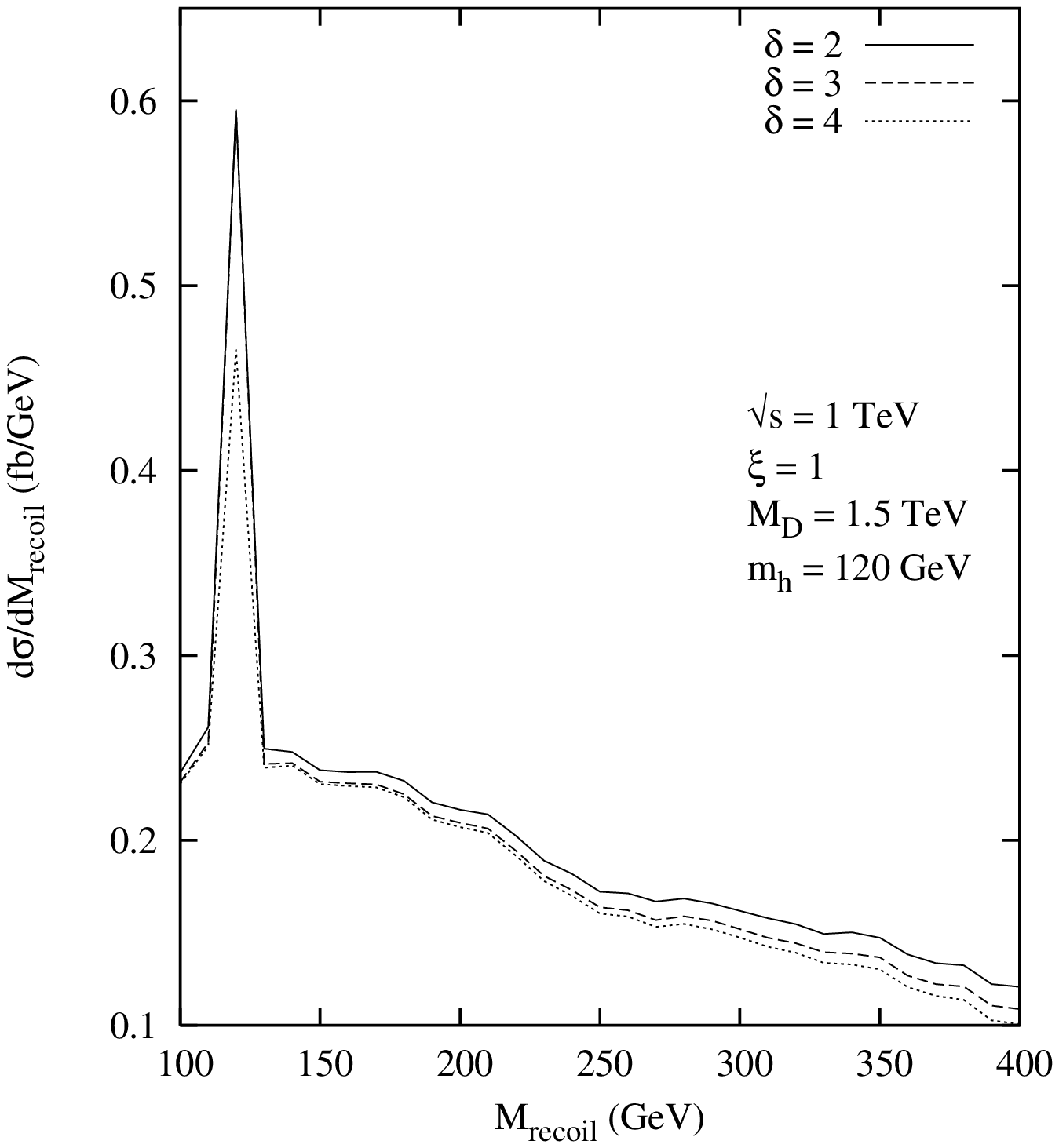}
\epsfxsize=8cm\epsfysize=9cm
                     \epsfbox{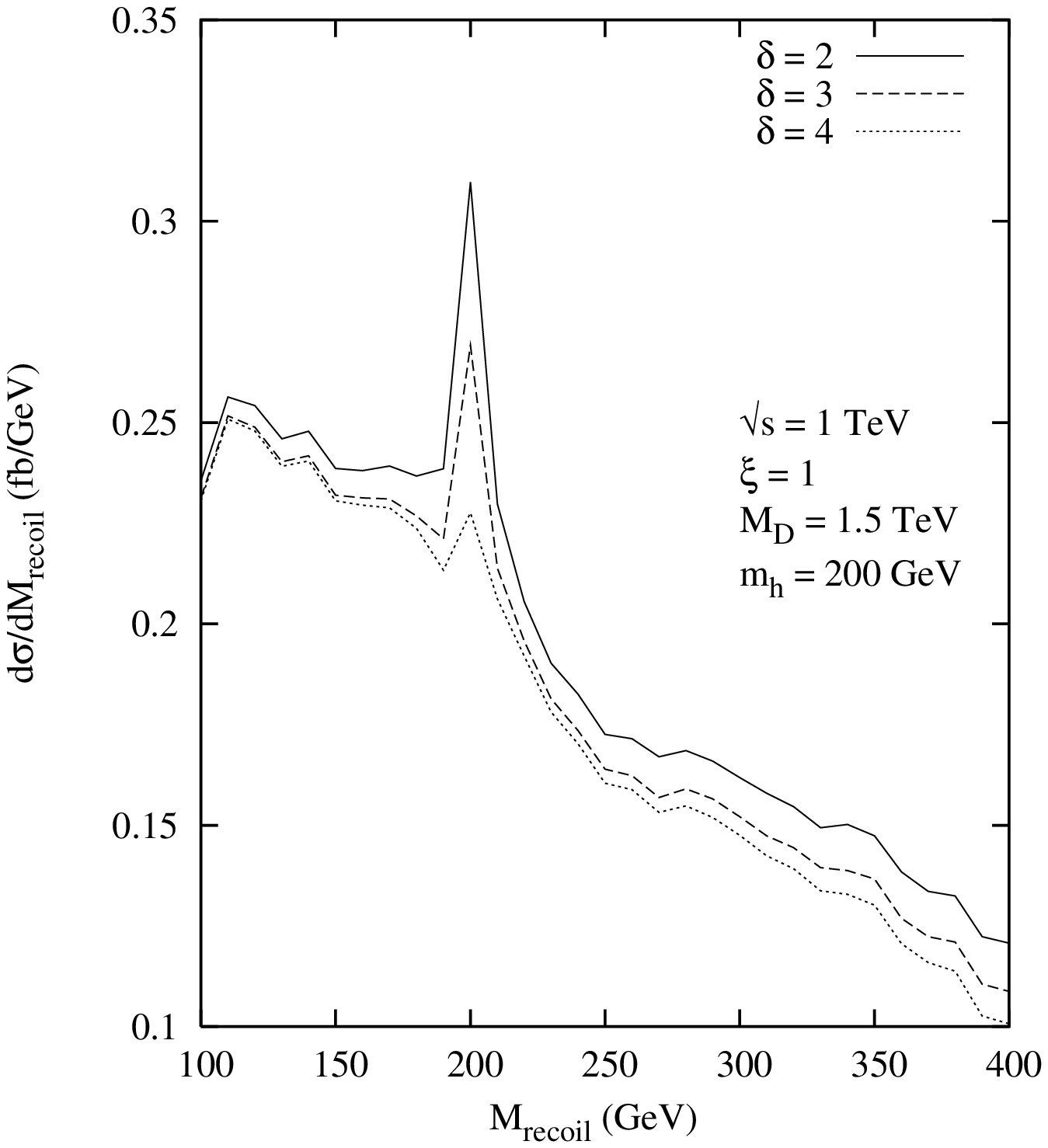}
}
\end{center}

\caption{\label{eecs}Recoil mass distribution for invisible Higgs 
for different values of $\delta$, superposed over the SM and 
gravitensor contributions, for $m_D$ = 1.5 TeV and  $\xi$ = 1.}
\end{figure}

The above conclusions are borne out
in Fig.~\ref{eecs} where we have plotted the recoil mass distribution (as
discussed above) for the $\mu^+ \mu^- ~ +~missing~energy$ events including
the  SM, gravitensor and invisibly decaying Higgs production.
The results are presented for 
three different values of the number of extra dimensions. 
In the two sets of plots, corresponding to Higgs 
masses 120 GeV (left) and 200 GeV (right),
respectively, the different degrees of visibility of the peak
is quite obvious, because of reasons discussed above.    
For the purpose of illustration, we have chosen $m_D = 1.5$
TeV, $\xi = 1$ and $\sqrt{s} = 1$ TeV. To take into account a
realistic detector resolution we have assumed a Gaussian spreading of
muon energy: $\Delta E /E = 0.15/\sqrt{E} + 0.01$
\cite{peskin}. The visibility of the peak for the lower Higgs mass 
is largely due to the angular cut. For a heavier Higgs, however,
reconstruction of the peak appears to be difficult.

The visibility of the proposed signal depends on the relative
rates of the invisible Higgs decay channel and
the SM and gravitensor continuum channels. The
standard procedure in such a situation is to compare the number of
Higgs signal events to the square root of the number of events coming
from the 'background' in the particular mass bin, in which we see the
peak. There are two caveats to this criteria. The number of background
(and signal) events should be sufficiently large so that square root
of this number represents 1$\sigma$ fluctuation of the background (in
Gaussian statistics). Secondly, we also must know the absolute normalization
of the background very well. While the first criterion can be
satisfied with high luminosity, the second one, in our case, poses a problem. 
There is a part of the number of events coming from gravitensor
production. This depends on unknown parameters like $\delta$ and $m_D$,
and thus is correlated with the number of signal (invisible Higgs)
events via these parameters.

In Table~\ref{tab1}, we present the cross-sections of $\mu^+ \mu^- +$
missing momentum final state from three different sources
separately. Instead of the total rates, the
cross-sections in a particular mass bin (of width 10 GeV) are 
more appropriate for our discussion and have been presented in
the tables. We choose two different Higgs masses: 120 GeV
and 200 GeV. 
Looking at the numbers in the table one can readily appreciate 
the smallness of gravitensor contributions to the relevant bins, 
in contrast  to the SM contributions.

Let us make some further comments about the numbers in the table. The
monotonic decrease of cross-sections (in a particular mass bin for a
fixed value of $m_D$ and $s_{ee}$) with increasing values of $\delta$
for invisible Higgs \footnote{The only exception
is when going from $\delta=2$ to $\delta=3$ for $m_h=120$ GeV,
$\sqrt{s}=1$ TeV, and $m_D=1.5$ TeV.  The slightly decreasing cross
section for this particular parameter set is due to the interplay of 
density of graviscalar states and a large portion of invisible Higgs 
partial width in the total width in Eq. (\ref{matel}).}
 and gravitensor is due to steady decrease of
available number density of graviscalar/tensor states as pointed out
earlier. Invisible Higgs cross-section also decreases with $e^+ e^-$
center-of-mass energy (compare any two corresponding entries for two
tables). This can be accounted by the s-channel propagators in the two
contributing diagrams (Fig.~\ref{diagrs}).  On the other hand, decrease of
graviton (spin-2) contribution to mass bins is modest with
center-of-mass energy. There are four graphs contributing to this spin-2
production (along with a $Z$-boson). The s-channel contribution goes down as
usual when we go to a higher center of mass energy. This fall
is somehow compensated by a four-point contact graph and we are left with 
a modest decrement in the cross-section with $e^+e^-$ center-of-mass energy. 
It is also to be noted that while the invisible Higgs contributions 
to the final state fall for higher center-of-mass energy, the gravitensor 
contribution remains practically unchanged. This is because the latter 
involves an integration over the entire tower that is kinematically available,
and a higher value of $\sqrt{s}$ allows one to integrate over a bigger tower.
For the invisible Higgs signal, however, this does not make any difference,
since the dominant contribution there comes from that portion of the
graviscalar tower which is close to the Higgs mass. Therefore,
the signature of scalar-graviscalar mixing stands out more prominently
for a linear collider operating at 500 GeV than one at 1 TeV.

\begin{table}
\begin{center}
\begin{tabular}{|c|c|c|c|c|c|c|c|}
\hline
\multicolumn{8}{|c|}{Cross-sections (fb) in bin of width 10 GeV;
$\sqrt{s_{ee}}$ = 0.5 TeV} \\
\hline \multicolumn{1}{|c|}{recoil} & \multicolumn{1}{|c}{}
& \multicolumn{3}{|c}{$m_D$ = 1.5 TeV} &
\multicolumn{3}{|c|}{$m_D$ = 3 TeV}  \\
 \cline{3-8} 
\multicolumn{1}{|c|}{mass bin} & \multicolumn{1}{|c}{} 
& \multicolumn{1}{|c}{$\delta$ = 2} &
\multicolumn{1}{|c|}{$\delta$ = 3} & \multicolumn{1}{|c}{$\delta$ = 4}
 & \multicolumn{1}{|c}{$\delta$ = 2} &
\multicolumn{1}{|c|}{$\delta$ = 3} & \multicolumn{1}{|c|}{$\delta$ = 4}\\ \hline
\multicolumn{1}{|r|}{} &\multicolumn{1}{|r|}{IH:}
& \multicolumn{1}{|c}{1.488} & \multicolumn{1}{|c|}{1.485} &
\multicolumn{1}{|c}{1.155} & \multicolumn{1}{|c|}{1.311}   &
\multicolumn{1}{|c}{0.592} & \multicolumn{1}{|c|}{0.075}  \\ 
\multicolumn{1}{|r|}{120 $\pm$ 5 }&\multicolumn{1}{|r|}{GT:}
& \multicolumn{1}{|c}{6.6$\times 10^{-3}$} & \multicolumn{1}{|c|}{0.001} &
\multicolumn{1}{|c}{0.0001} & \multicolumn{1}{|c|}{4.4$\times 10^{-4}$} &
\multicolumn{1}{|c}{3.3 $\times10^{-5}$} & 
\multicolumn{1}{|c|}{2 $\times10^{-6}$}  \\ 
\multicolumn{1}{|c|}{GeV}&\multicolumn{1}{|r|}{SM:}
& \multicolumn{1}{|c}{1.95} & \multicolumn{1}{|c|}{1.95} &
\multicolumn{1}{|c}{1.95} & \multicolumn{1}{|c|}{1.95}   &
\multicolumn{1}{|c}{1.95} & \multicolumn{1}{|c|}{1.95}  \\ 

\hline
\multicolumn{1}{|r|}{}&\multicolumn{1}{|r|}{IH:}
& \multicolumn{1}{|c}{0.321} & \multicolumn{1}{|c|}{0.205} &
\multicolumn{1}{|c}{0.070} & \multicolumn{1}{|c|}{0.041}  &
\multicolumn{1}{|c}{0.009} & \multicolumn{1}{|c|}{0.001}   \\ 
\multicolumn{1}{|r|}{200 $\pm 5$}& \multicolumn{1}{|r|}{GT:}
& \multicolumn{1}{|c}{0.013} & \multicolumn{1}{|c|}{0.0036} &
\multicolumn{1}{|c}{7.4 $\times10^{-4}$} & 
\multicolumn{1}{|c|}{8.3$\times 10^{-4}$} &
\multicolumn{1}{|c}{1.1 $\times10^{-4}$} 
& \multicolumn{1}{|c|}{1.1 $\times10^{-5}$}  \\ 
\multicolumn{1}{|c|}{GeV} & \multicolumn{1}{|r|}{SM:}
& \multicolumn{1}{|c}{0.873} & \multicolumn{1}{|c|}{0.873} &
\multicolumn{1}{|c}{0.873} & \multicolumn{1}{|c|}{0.873}   &
\multicolumn{1}{|c}{0.873} & \multicolumn{1}{|c|}{0.873}  \\ 

\hline
\multicolumn{1}{|r|}{}&\multicolumn{1}{|r|}{IH:}
& \multicolumn{1}{|c}{0.067} & \multicolumn{1}{|c|}{0.059} &
\multicolumn{1}{|c}{0.035} & \multicolumn{1}{|c|}{0.011}  &
\multicolumn{1}{|c}{0.0037} & \multicolumn{1}{|c|}{0.00075}   \\ 
\multicolumn{1}{|r|}{300 $\pm 5$}& \multicolumn{1}{|r|}{GT:}
& \multicolumn{1}{|c}{0.0311} & \multicolumn{1}{|c|}{0.0125} &
\multicolumn{1}{|c}{0.0039} & 
\multicolumn{1}{|c|}{1.9$\times 10^{-3}$} &
\multicolumn{1}{|c}{3.9 $\times10^{-4}$} 
& \multicolumn{1}{|c|}{6.1 $\times10^{-5}$}  \\ 
\multicolumn{1}{|c|}{GeV} & \multicolumn{1}{|r|}{SM:}
& \multicolumn{1}{|c}{0.297} & \multicolumn{1}{|c|}{0.297} &
\multicolumn{1}{|c}{0.297} & \multicolumn{1}{|c|}{0.297}   &
\multicolumn{1}{|c}{0.297} & \multicolumn{1}{|c|}{0.297}  \\ 
\hline
\end{tabular}
\end{center}
\begin{center}
\begin{tabular}{|c|c|c|c|c|c|c|c|}
\hline
\multicolumn{8}{|c|}{Cross-sections (fb) in bin of width 10 GeV;
$\sqrt{s_{ee}}$ = 1 TeV} \\
\hline  \multicolumn{1}{|c|}{recoil} & \multicolumn{1}{|c}{}
& \multicolumn{3}{|c}{$m_D$ = 1.5 TeV} &
\multicolumn{3}{|c|}{$m_D$ = 3 TeV}  \\
 \cline{3-8}\multicolumn{1}{|c|}{mass bin} & \multicolumn{1}{|c}{} 
& \multicolumn{1}{|c}{$\delta$ = 2} &
\multicolumn{1}{|c|}{$\delta$ = 3} & \multicolumn{1}{|c}{$\delta$ = 4}
 & \multicolumn{1}{|c}{$\delta$ = 2} &
\multicolumn{1}{|c|}{$\delta$ = 3} & \multicolumn{1}{|c|}{$\delta$ = 4}\\ \hline
\multicolumn{1}{|r|}{} &\multicolumn{1}{|r|}{IH:}
& \multicolumn{1}{|c}{0.341} & \multicolumn{1}{|c|}{0.346} &
\multicolumn{1}{|c}{0.217} & \multicolumn{1}{|c|}{0.219}   &
\multicolumn{1}{|c}{0.093} & \multicolumn{1}{|c|}{0.011}  \\ 
\multicolumn{1}{|r|}{120 $\pm$ 5} &\multicolumn{1}{|c|}{GT:}
& \multicolumn{1}{|c}{0.0058} & \multicolumn{1}{|c|}{9.2$\times10^{-4}$ } &
\multicolumn{1}{|c}{1.1$\times10^{-4}$} & \multicolumn{1}{|c|}{3.6$\times 10^{-4}$} &
\multicolumn{1}{|c}{2.9 $\times10^{-5}$} & \multicolumn{1}{|c|}{1.8 $\times10^{-6}$}  \\ 
\multicolumn{1}{|c|}{GeV} &\multicolumn{1}{|c|}{SM:}
& \multicolumn{1}{|c}{0.247} & \multicolumn{1}{|c|}{0.247} &
\multicolumn{1}{|c}{0.247} & \multicolumn{1}{|c|}{0.247}   &
\multicolumn{1}{|c}{0.247} & \multicolumn{1}{|c|}{0.247}  \\ 

\hline
\multicolumn{1}{|r|}{} & \multicolumn{1}{|r|}{IH:}
& \multicolumn{1}{|c}{0.093} & 
\multicolumn{1}{|c|}{0.059} &
\multicolumn{1}{|c}{0.020} & \multicolumn{1}{|c|}{0.012}  &
\multicolumn{1}{|c}{0.0027} & \multicolumn{1}{|c|}{0.0003}   \\ 
\multicolumn{1}{|r|}{200 $\pm$ 5} &\multicolumn{1}{|c|}{GT:}
& \multicolumn{1}{|c}{9.8$\times10^{-3}$} & \multicolumn{1}{|c|}{0.0028} &
\multicolumn{1}{|c}{0.0005} & \multicolumn{1}{|c|}{6.5$\times 10^{-4}$} &
\multicolumn{1}{|c}{8.7 $\times10^{-5}$} & \multicolumn{1}{|c|}{8.2 $\times10^{-6}$}  \\ 
\multicolumn{1}{|c|}{GeV} &\multicolumn{1}{|c|}{SM:}
& \multicolumn{1}{|c}{0.206} & \multicolumn{1}{|c|}{0.206} &
\multicolumn{1}{|c}{0.206} & \multicolumn{1}{|c|}{0.206}   &
\multicolumn{1}{|c}{0.206} & \multicolumn{1}{|c|}{0.206}  \\ 

\hline
\multicolumn{1}{|r|}{}&\multicolumn{1}{|r|}{IH:}
& \multicolumn{1}{|c}{0.035} & \multicolumn{1}{|c|}{0.03} &
\multicolumn{1}{|c}{0.018} & \multicolumn{1}{|c|}{0.006}  &
\multicolumn{1}{|c}{0.002} & \multicolumn{1}{|c|}{0.00038}   \\ 
\multicolumn{1}{|r|}{300 $\pm 5$}& \multicolumn{1}{|r|}{GT:}
& \multicolumn{1}{|c}{0.016} & \multicolumn{1}{|c|}{0.0065} &
\multicolumn{1}{|c}{0.0021} & 
\multicolumn{1}{|c|}{1.1$\times 10^{-3}$} &
\multicolumn{1}{|c}{2.0 $\times10^{-4}$} 
& \multicolumn{1}{|c|}{3.4 $\times10^{-5}$}  \\ 
\multicolumn{1}{|c|}{GeV} & \multicolumn{1}{|r|}{SM:}
& \multicolumn{1}{|c}{0.145} & \multicolumn{1}{|c|}{0.145} &
\multicolumn{1}{|c}{0.145} & \multicolumn{1}{|c|}{0.145}   &
\multicolumn{1}{|c}{0.145} & \multicolumn{1}{|c|}{0.145}  \\ 
\hline
\end{tabular}
\end{center}

\caption{\label{tab1}Cross-sections in the relevant bins 
of recoil mass (of the system recoiling against $\mu^+ \mu^-$ pair)
for invisible Higgs (IH), gravitensor (GT) and the
Standard Model (SM) for $e^+e^-$ center of mass energy of 0.5 and 1
TeV. Entries in the left-most columns of each table indicate the 
central values and widths of the mass bins, we are interested. For 
IH, the central value of the recoil mass bin indicates the Higgs 
mass. }

\end{table}

\section{Summary and conclusions}
We have investigated the detection prospects of a Higgs boson that
has mixing with a quasi-continuous tower of graviscalars in a
scenario with large compact extra dimensions. This causes the Higgs
to develop an invisible decay width. In the context of a linear $e^+ e^-$
collider, however, such invisibility brings in additional problems
in reconstructing the Higgs boson as a recoil mass peak against an
identified Higgs boson. This is because of (a) the simultaneous presence
of graviton continuum production in association with a $Z$-boson, and (b) the
broadening of the Higgs peak due to enhancement of the total effective
decay width. We find that while a judicious angular cut partially
alleviated the difficulty, the broadening of the peak remains a problem,
particularly for $m_H > 2 m_W$.  Therefore, the search for a Higgs
boson in a scenario of this kind, and more importantly, the detailed
investigation of its properties, has wider ramification than an
invisibly decaying Higgs envisioned in most other models.  

When this work was nearing its completion, a corresponding study in 
the context of  LHC \cite{BDGW} appeared in the literature, where an
attempt has been made to point out the detectability of  
a light  invisible Higgs in the ADD-model  hadron colliders in the
gauge boson fusion channel.  However, a complete answer is not yet
available on whether continuum graviton production can be a problem here, 
too, since such production can take place even in gluon fusion. Therefore,
our view is that a full estimate of direct gravitensor and graviscalar 
production rates giving rise to the same final states is required before 
a conclusion can be reached on the signature Higgs-graviscalar mixing. 
We are also aware of \cite{ms},
where invisible Higgs in linear colliders has been considered
for $Z$ decaying to a quark pair. Again, the difficulties arising from
direct graviton  production need to be addressed before reaching a verdict
in such a study.

\vspace*{1cm}
\noindent
{\Large\bf Acknowledgments}

\vspace*{0.5cm}
\noindent
AD, KH and JL thank the Academy of Finland
(project numbers 48787, 104368, and 54023) for financial support.
BM wishes to acknowledge the hospitality of Katri Huitu and the Helsinki
Institute of Physics at the time when this work was initiated.

\end{document}